\documentclass{PoS}

\usepackage{wrapfig}

\usepackage{tikz}
\definecolor{valkoinen}{rgb}{1 1 1}
\definecolor{hysininen}{rgb}{0 0 0}
\definecolor{hyharmaa}{rgb}{0.5 0.5 0.5}
\definecolor{mloranssi}{rgb}{1 1 1}
\definecolor{vaaleaoranssi}{rgb}{1 1 1}
\definecolor{vaalvihrea}{rgb}{0.2 0.9 0.1}
\definecolor{omaviol}{rgb}{0.7 0.1 0.7}
\definecolor{omavihrea}{rgb}{0.1 0.3 0.1}
\definecolor{omapun}{rgb}{0.5 0.0 0.15}
\definecolor{omasin}{rgb}{0.0 0.15 0.5}
\definecolor{omaor}{rgb}{0.45 0.3 0}

\newcommand{\ud}{\, \mathrm{d}}
\newcommand{\nc}{{N_\mathrm{c}}}
\newcommand{\nf}{{N_\mathrm{f}}}

\newcommand{\as}{\alpha_{\mathrm{s}}}
\newcommand{\lqcd}{\Lambda_{\mathrm{QCD}}}
\newcommand{\tr}{\, \mathrm{Tr} \, }

\newcommand{\nr}[1]{(\ref{#1})}

\newcommand{\qso}{Q_{\textnormal{s0}}}
\newcommand{\qs}{Q_{\textnormal{s}}}

\newcommand{\fig}{Fig.~}

\usepackage{amssymb}
\usepackage{amsmath}

\title{Solving the NLO BK equation in coordinate space}
\ShortTitle{Solving the NLO BK equation in coordinate space}
\author{\speaker{T. Lappi}\\
Department of Physics, 
 P.O. Box 35, 40014 University of Jyv\"askyl\"a, Finland
\\
Helsinki Institute of Physics, P.O. Box 64, 00014 University of Helsinki,
Finland \\
E-mail: \email{tuomas.v.v.lappi@jyu.fi}}
\author{H. M\"antysaari\\
Department of Physics, 
 P.O. Box 35, 40014 University of Jyv\"askyl\"a, Finland \\
E-mail: \email{heikki.mantysaari@jyu.fi}}
\abstract{
We present results from a numerical solution of the next-to-leading order (NLO) Balitsky-Kovchegov (BK) equation in coordinate space in the large $\nc$ limit. We show that the solution is not stable for initial conditions that are close to those used in phenomenological applications of the leading order equation. We identify the problematic terms in the NLO kernel as being related to large logarithms of a small parent dipole size, and also show that rewriting the equation in terms of the ``conformal dipole'' does not remove the problem. Our results qualitatively agree with expectations based on the behavior of the linear NLO BFKL equation.
}

\FullConference{XXIII International Workshop on Deep-Inelastic 
Scattering\\
                 27 April - May 1 2015\\
                 Dallas, Texas}

\begin{document}
\section{Introduction}
\begin{wrapfigure}{r}{0.3\textwidth}
\centerline{
\begin{tikzpicture}[thick,scale=1.2]
\node (x) at (0,0) {$x$};
\node (zp) at (0,2) {$z'$};
\node (z) at (2,0.5) {$z$};
\node (y) at (2.5,2.5) {$y$};
\path[line width=1pt,<->] (z) edge (zp);
\draw[line width=1pt,<->,omaviol] (x) -- node[above,pos=0.7] {$r$} (y);
\draw[line width=1pt,<->,omaor] (x) -- node[left] {$X'$} (zp);
\draw[line width=1pt,<->,omavihrea] (y) -- node[above] {$Y'$} (zp);
\draw[line width=1pt,<->,omapun] (x) -- node[below] {$X$} (z);
\draw[line width=1pt,<->,omasin] (y) -- node[right] {$Y$} (z);
\end{tikzpicture}
}
\caption{Coordinates for the eikonal Wilson lines and their separations}
\label{fig:coords}
\end{wrapfigure}
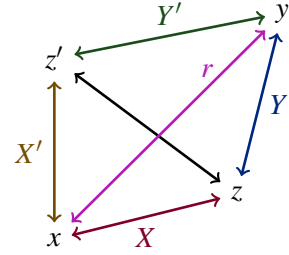
The CGC picture~\cite{Gelis:2010nm} of high energy QCD has been 
succesfully applied to describe deep inelastic scattering, single particle 
production 
 and particle correlations 
in forward rapidity proton-proton and proton nucleus collisions. It also forms a basis for understanding the initial stage of deconfined QCD matter created in ultrarelativistic nucleus-nucleus collisions. An essential ingredient in these calculations are the JIMWLK equation and its mean field limit --- the Balitsky-Kovchegov (BK) equation~\cite{Balitsky:1995ub,Kovchegov:1999yj}. They describe the energy, or equivalently Bjorken-$x$, dependence of correlators of Wilson lines in the target color field. Cross sections for different processes are then expressed in terms of these Wilson line correlators. 

Recently first steps have been taken to develop this phenomenological picture to next-to-leading order (NLO) accuracy, both for the evolution equations themselves~\cite{Balitsky:2008zza,Balitsky:2013fea,Kovner:2013ona} and for specific scattering processes~\cite{Balitsky:2010ze,Beuf:2011xd,Chirilli:2011km,Altinoluk:2014eka}. 
There remain both conceptual and numerical challenges in carrying out the full CGC phenomenology program at NLO accuracy.
This talk discusses in more detail one of these aspects, by presenting the result~\cite{Lappi:2015fma} from a direct ``brute force'' numerical solution of the NLO BK equation as it is written down in Ref.~\cite{Balitsky:2008zza}.

\section{The equation}

In the large-$\nc$ limit (but assuming $\nf \sim \nc$)
 we can write the evolution equation derived in Ref.~\cite{Balitsky:2008zza}
for the dipole as:
\begin{multline}
\label{eq:nlobk}
	\partial_y S(r) = \frac{\as \nc}{2\pi^2} K_1 \otimes [S(X)S(Y)-S(r)] 
		+ \frac{\as^2 \nf \nc}{8\pi^4} K_f \otimes S(Y)[S(X')-S(X)]
\\
		+ \frac{\as^2 \nc^2}{8\pi^4} K_2 \otimes 
[S(X)S(z-z')S(Y')-S(X)S(Y)], \quad \textnormal{ with }  S(r) = \frac{1}{\nc} 
\left< \tr U^\dag(x)U(y) \right>,
\end{multline}
Here $U(x)$ is a fundamental representation Wilson line describing the
propagation of a high energy probe through the dense color field of the target.
The Wilson lines are needed at coordinates $x,y,z,z'$
in the two-dimensional transverse plane, with the
six distances between them denoted as $r,X,X',Y,Y'$ and $z-z'$ as shown in 
\fig\ref{fig:coords}. The convolutions $\otimes$ denote integrations over $z$
 (in $K_1$) or $z$ and $z'$ (the other terms). We replace the terms explicitly 
proportional to the $\beta$-function coefficient in the first kernel
$K_1$ by the ``Balitsky'' 
running coupling prescription~\cite{Balitsky:2006wa}. With this replacement the
 explicit expression for the first kernel is
\begin{multline}
\frac{\as \nc}{2\pi^2} K_1 = \frac{\as(r) \nc}{2\pi^2} \left[\frac{r^2}{X^2Y^2} + \frac{1}{X^2} \left(\frac{\as(X)}{\as(Y)}-1\right) + \frac{1}{Y^2} \left(\frac{\as(Y)}{\as(X)}-1\right) \right] 
\\
		+ \frac{\as(r)^2 \nc^2}{8\pi^3} \frac{r^2}{X^2Y^2} \left[ \frac{67}{9} - \frac{\pi^2}{3} - \frac{10}{9} \frac{\nf}{\nc} - 2\ln \frac{X^2}{r^2} \ln \frac{Y^2}{r^2} \right].
\end{multline}
The first term inside first the square bracket in $K_1$ is the leading order 
fixed coupling kernel. The whole first square bracket in $K_1$ corresponds
to the ``Balitsky'' running coupling prescription often used in LO 
phenomenology~\cite{Albacete:2010bs,Lappi:2013zma}, with the second square 
bracket being a pure NLO term.
The coupling constant in front of the purely NLO kernels $K_2$ and $K_f$ 
is taken to depend on the parent dipole size $r$ and the explicit expressions for the kernels are
\begin{align}
K_2 &= -\frac{2}{(z-z')^4} + \bigg[ \frac{X^2 Y'^2 + X'^2Y^2 - 4r^2(z-z')^2}{(z-z')^4(X^2Y'^2 - X'^2Y^2)} 
\\ & \quad \quad 
+ \frac{r^4}{X^2Y'^2(X^2Y'^2 - X'^2Y^2)} + \frac{r^2}{X^2Y'^2(z-z')^2} \bigg]
 \ln \frac{X^2Y'^2}{X'^2Y^2} \nonumber
\\
 K_f &= \frac{2}{(z-z')^4}  
	- \frac{X'^2Y^2 + Y'^2 X^2 - r^2 (z-z')^2}{(z-z')^4(X^2Y'^2 - X'^2Y^2)}  \ln \frac{X^2Y'^2}{X'^2Y^2} .
\end{align}

In addition to the 
$\beta$-function terms, two kinds of logarithms appear in the kernels. The ones
in $K_2$ and $K_f$ depend on conformal ratios of four distances, and vanish
in the limit $r \to 0$. The first kernel $K_1$, on the other hand, has a 
nonconformal double logarithm that diverges in the limit $r\to 0$. Although
this is an integrable singularity in the $z$-integral, it nevertheless 
has a problematic effect on the evolution equation, as we will show 
in the following. As an initial condition we use a parametrization 
\begin{equation}\label{eq:initc}
  N(r) \equiv 1-S(r) = 
1 - \exp \left[ -\frac{(r^2 \qso^2)^\gamma}{4} 
\ln \left(\frac{1}{r \lqcd}+ e\right)\right],
\end{equation}
with two tunable parameters:    
\begin{itemize}
\item The ratio $\qso/\lqcd$  essentially determines value of $\as$
and controls the overall relative importance of the NLO corrections.
\item The anomalous dimension $\gamma$ controls the shape of the initial 
condition. Leading order fits to HERA data using the parametrization 
\nr{eq:initc} prefer a value $\gamma\gtrsim 1$ which then becomes 
$\gamma\sim 0.8$ during the BK evolution. Since in an NLO fit also the 
impact factor relating the cross section and the dipole amplitude
$N(r)$ should be different from the LO one, without performing the
full fit it is not a priori obvious 
what would be a value favored by experimental data.
\end{itemize}

\section{Properties of the solution}

\begin{figure}
\includegraphics[width=0.4885\textwidth]{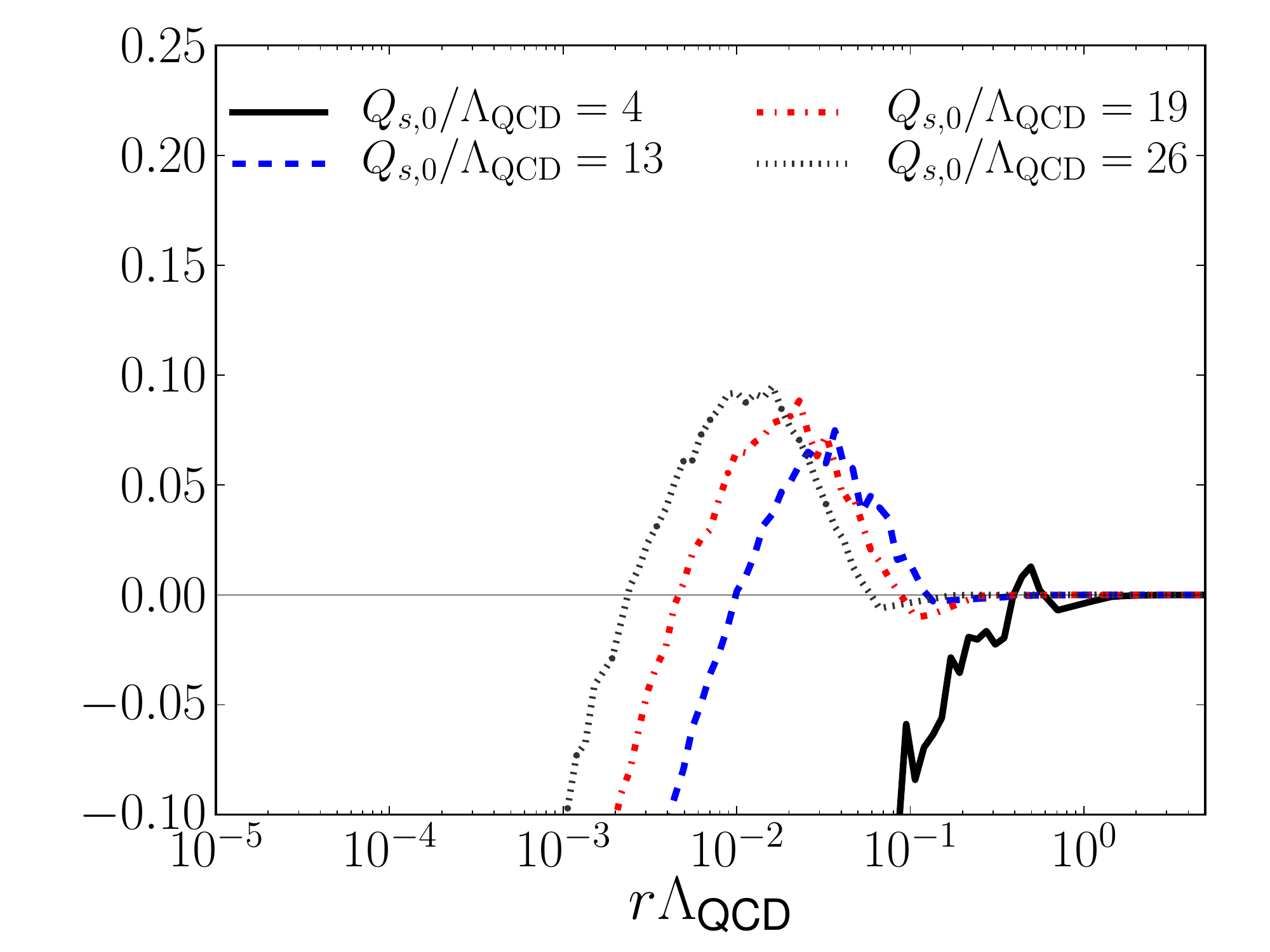}
\raisebox{1.5pt}{\includegraphics[width=0.5115\textwidth]{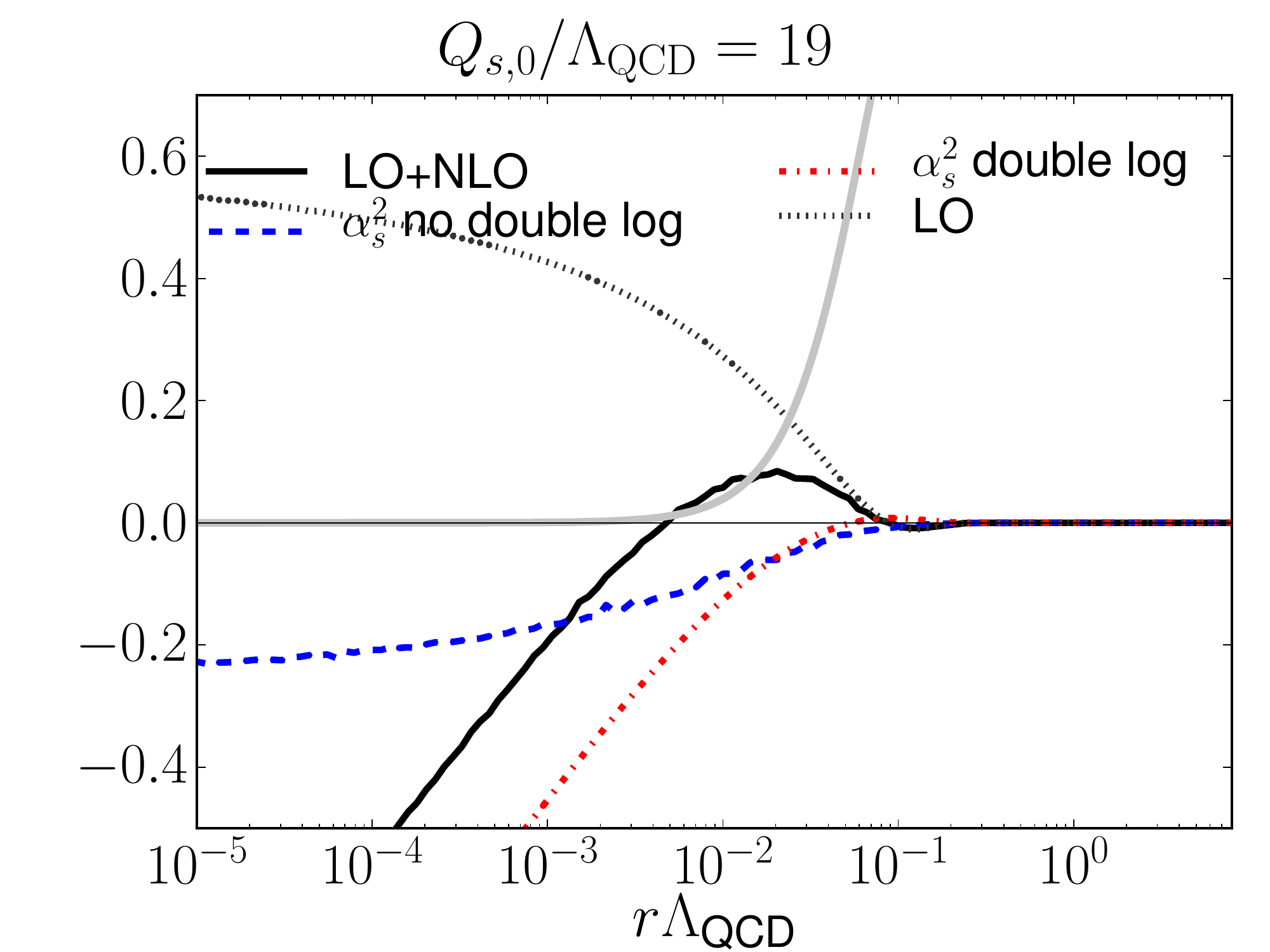}}
\caption{Left: Evolution speed $\partial_y N(r)/N(r)$ at the initial condition 
$y=0$ for the MV model initial condition, $\gamma=1$
Right: 
Contribution of different parts of the equation \protect\nr{eq:nlobk} to 
the evolution speed $\partial_y N(r)/N(r)$. Also the dipole amplitude $N(r)$ is 
shown with a thick grey line.
}
\label{fig:dndy_mv} \label{fig:dndy_contribs}
\end{figure}

Figure~\ref{fig:dndy_mv} (left) shows the logarithmic evolution speed 
$\partial_y N(r)/N(r)$ for the MV model initial condition $\gamma=1$. 
Firstly it is obvious that for small values of $\qs/\lqcd$, i.e. effectively
large couplings, the the evolution speed is negative at all values if $r$. 
This means that the NLO corrections are large and negative, and the scattering
amplitude actually decreases with energy; a very nonintuitive result. For
smaller typical values of $\as$ the region around the ``front'' 
$r\sim 1/\qs$ behaves in a reasonable way, but for very small dipoles
the evolution speed still diverges as $\partial_y N / N \sim \ln r$.
Figure~\ref{fig:dndy_contribs} (right) shows the contributions from 
different parts of the equation. The LO term gives a positive
$\partial_y N / N$ that approaches a constant at $r\to 0$. The
divergence for small $r$ is due to the nonconformal double logarithm, while
the other NLO corrections yield a contribution that is negative, but smaller 
in magnitude than the leading order result.

While having $\partial_y N(r) <0 $ or $N(r)< 0$ is not very physical, it is
not in itself a mathematical contradiction. A divergent 
$\partial_y N / N \sim \ln r$ in the limit $r\to 0$ is, however, a signal 
of an instability in the system. One way to see this is
 the following simple argument. Let us consider a small but finite 
interval in rapidity, $\ud y$, as in a numerical solution of the evolution 
equation. A diverging $\partial_y N / N$ for $r\to 0$ means that there is a
small but finite $r$ below which $N$ becomes  negative already in this one
step in rapidity. This immediately makes the equation 
unstable, since the convergence of the $z$-integral of the leading order 
equation requires $N(r)\to 0$ for $r\to 0$. A finite $N(r=0)$ is also 
inconsistent with the definition  $N(x-y) = 1-\frac{1}{\nc}\tr U^\dag(x)U(y)$ 
in terms of Wilson lines. To avert this problem in the numerics, 
we always   enforce $N(r)\geq 0$ by hand.

\section{The anomalous dimension}

\begin{figure}
\includegraphics[width=0.33\textwidth]{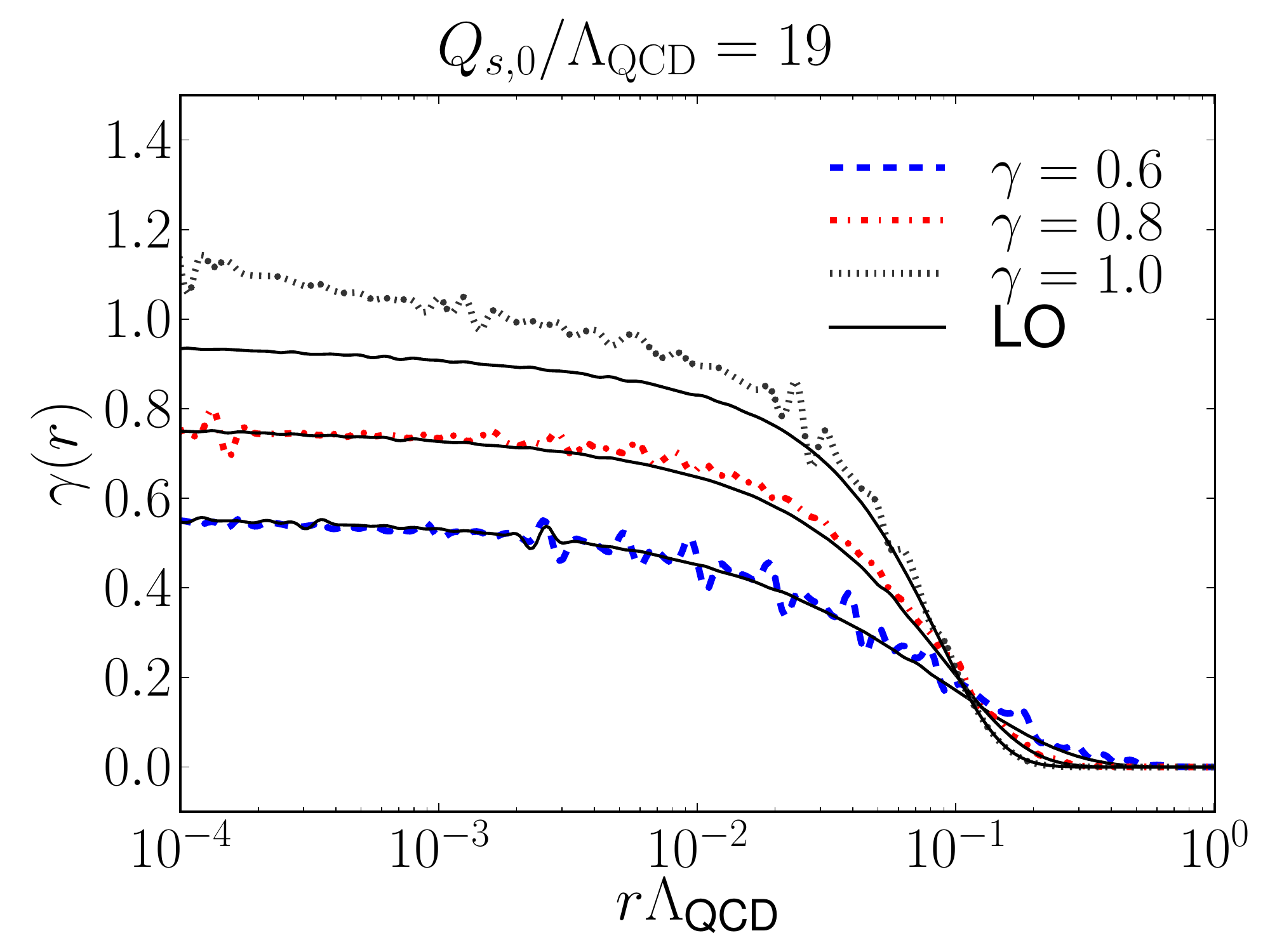}
\includegraphics[width=0.33\textwidth]{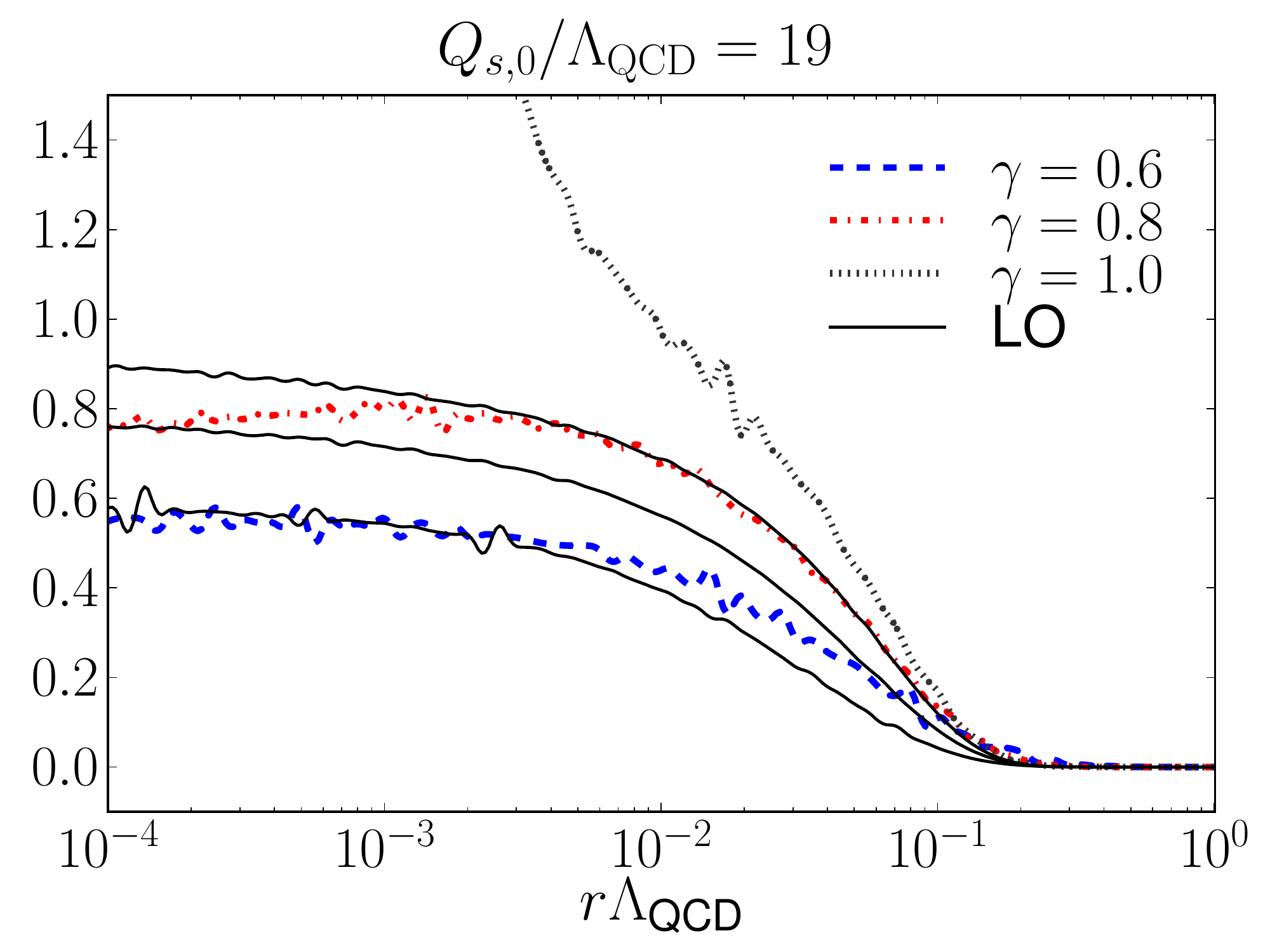}
\includegraphics[width=0.33\textwidth]{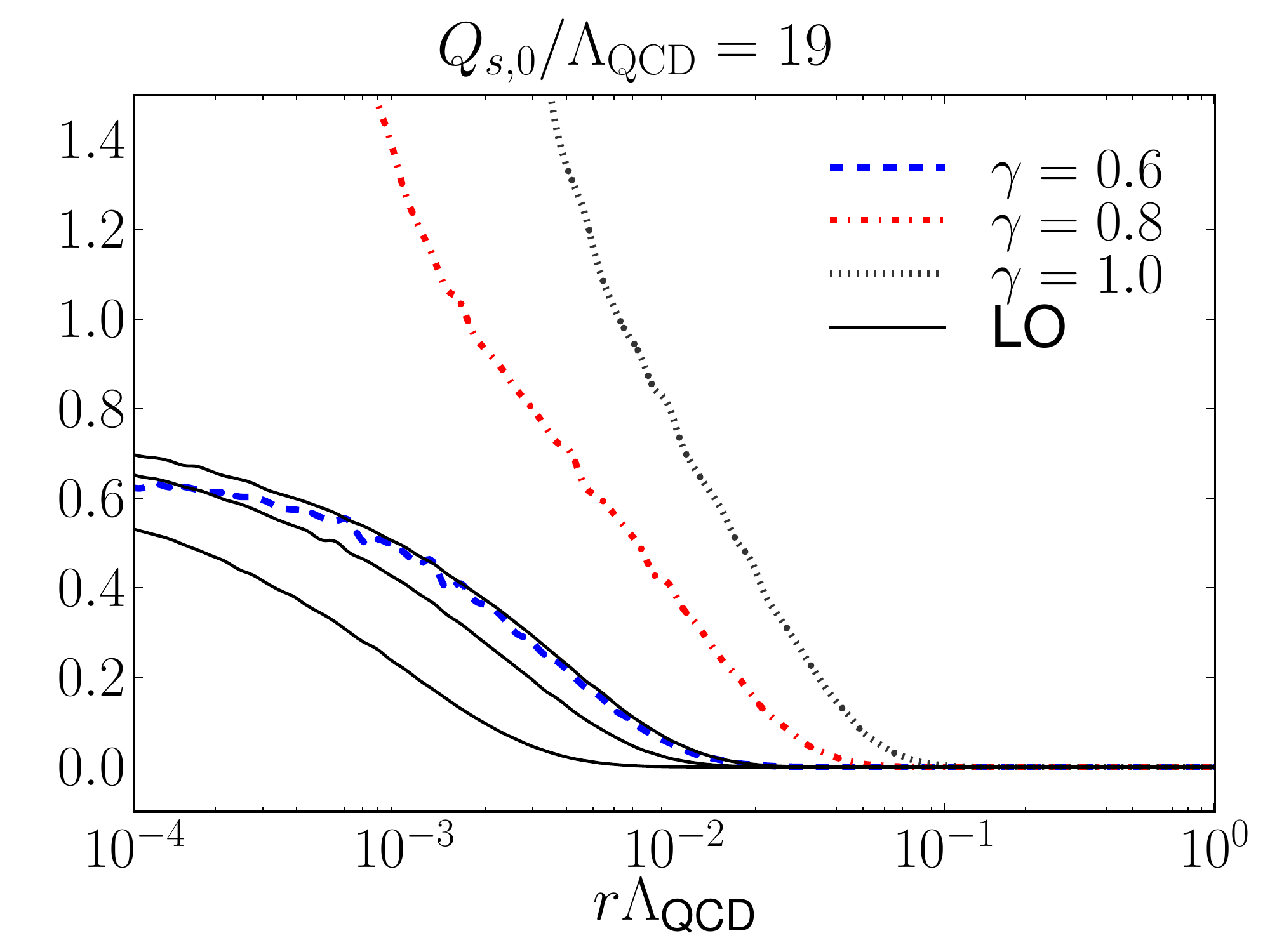}
\caption{Anomalous dimension $\gamma(r) =  \ud \ln N(r) /\ud \ln r^2$ at $y=1,\, 5,\, 30$.}
\label{fig:gamma}
\end{figure}

Some more insight into the behavior of the equation can be obtained by 
following the development of the $r$-dependent anomalous dimension 
that we define here as 
\begin{equation}
\gamma(r) =  \frac{\ud \ln N(r) }{\ud \ln r^2} .
\end{equation}
In terms of the anomalous dimension
we can discuss the behavior of the equation at small $r$ by 
parametrizing the amplitude as $N(r)\sim (\qs r)^{2\gamma}$. 
For an appropriate $\gamma$ the leading order equation maintains this
form, with $\qs^2\sim e^{\lambda y}$ and  
$\partial_y N/N \approx 2 \gamma \lambda >0$. 
If, due to large NLO corrections, one has
$\partial_y N/N \to - c < 0$ for $r\to 0$, this means that the solution
behaves as $N \sim e^{-c y}$: mathematically  this is not problematic, 
but physically it is unnatural to have the scattering  amplitude 
decrease with energy. If, however, $\partial_y N / N \sim c \ln r$, we
can parametrize  $N(r) \sim (\qs r)^{2\gamma(y)}$, with $\gamma(y) \sim y$. 
In other words, for a diverging evolution speed the functional form of
the amplitude as a function of $r$ gets steeper with the evolution. 
If one enforces $N>0$ for small $r$ and $N<1$ for $r\to \infty$,
 the amplitude eventually develops into a 
discontinuous form  $N(r) \sim \theta(r-1/\qs)$.
The evolution of $\gamma(r)$ for different initial conditions is shown in 
\fig\ref{fig:gamma}, where this behavior can be clearly seen.
 The unstable nature of the equation shows up
as an increase in $\gamma(r)$ for small $r$. This increase starts immediately 
for the initial condition with $\gamma=1$, more  slowly for the initial condition
$\gamma=0.8$ and is not noticeable within the rapidity range studied 
here for $\gamma=0.6$.

\section{The composite conformal dipole}

In Ref.~\cite{Balitsky:2009xg} Balitsky and Chirilli note that the 
nonconformal double logarithmic term appears also in the equation for
the conformal $N=4$ Super Yang-Mills theory. It is therefore
interpreted as an artefact of a cutoff that breaks conformal invariance. 
The authors then propose to correct for this effect order by order 
in perturbation theory by introducing a ``composite conformal dipole'' operator
defined as
\begin{equation}\label{eq:confs}
S(r)^\text{conf} = S(r) - \frac{\as \nc}{4\pi^2} \int \ud^2 z \frac{r^2}{X^2Y^2} \ln \frac{a r^2}{X^2Y^2} [ S(X)S(Y) - S(r) ],
\end{equation}
\begin{wrapfigure}{r}{0.5\textwidth}
\includegraphics[width=0.5\textwidth]{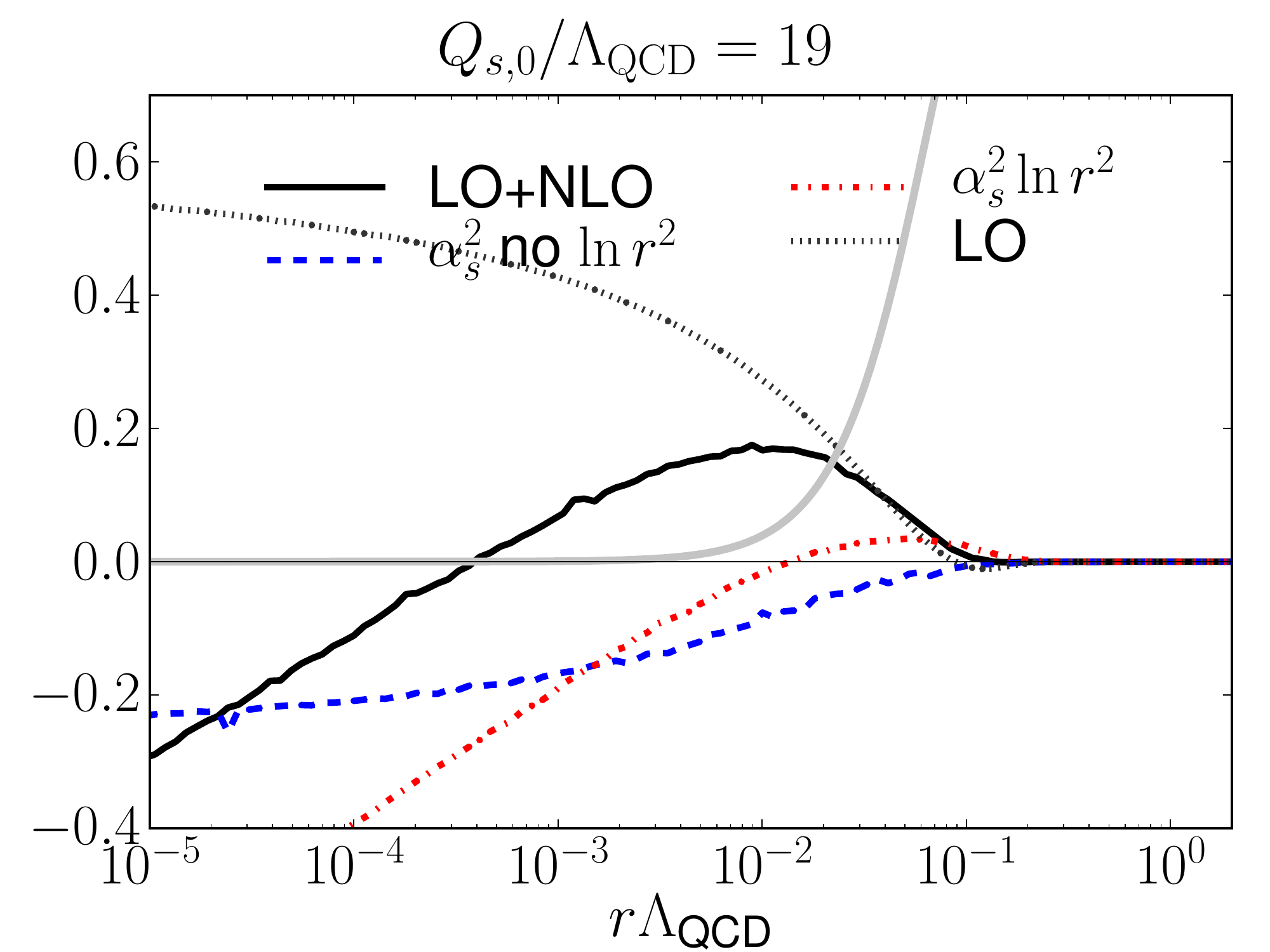}
\caption{Contribution of different terms to the evolution speed
$\partial_y N/N$
of the conformal dipole.}
\label{fig:confs}
\end{wrapfigure}
with $a$ a dimensionful constant that drops out of the final equation.
Rewriting the equation \nr{eq:nlobk} in terms of the conformal 
dipole~\nr{eq:confs} removes the nonconformal double logaritm, but
introduces an additional term in the other kernel $K_2$ that behaves as
$\ln r^2$ for small $r$. We have checked numerically that 
the qualitative behavior of the conformal dipole equation
 remains the same as the original one. 
As shown in \fig\ref{fig:confs}, it is now this new $\ln r^2$ term
which is responsible for the leading behavior at small $r$.

In conclusion, we have performed the first
numerical solution of the full NLO BK equation directly in coordinate space. 
The NLO corrections are negative, implying  a slower energy dependence of
cross sections than with the LO equation. This generically leads to a 
better agreement with experimental data. The equation has, however, a 
double logaritmic term that causes a problematic behavior for small
dipoles, i.e. high $Q^2$. It seems evident that these large logarithms 
will need to be resummed in order for the equation to be useful in 
practical phenomenological work.

\paragraph{Acknowledgements}
 This work has been supported by the Academy of Finland, projects 
267321 and 273464, the Graduate School of Particle and Nuclear Physics (H.M.)
and by computing resources from
CSC -- IT Center for Science in Espoo, Finland.

\bigskip

\bibliography{spires}
\bibliographystyle{JHEP-2modlong}

\end{document}